# Measuring Minority Carrier Diffusion Length Using High-Injection Scanning Photocurrent Microscopy

XIUJUN LIAN[1] and HANWEI GAO[1,2,3]*

1. Department of Physics, Florida State University, Tallahassee, FL 32306, USA

2. Materials Science and Engineering Program, Florida State University, Tallahassee, FL 32306, USA

3. Condensed Matter Science, National High Magnetic Field Laboratory, Tallahassee, FL 32310, USA

ABSTRACT Scanning photocurrent microscopy (SPCM) has been widely used for characterizing charge transport properties, in particular, the minority carrier diffusion length of semiconductors. However, studying lightly doped or intrinsic semiconductors using SPCM remained challenging. Methods used in previous work required low levels of optical injection, which could not be fulfilled easily in semiconductors with lower carrier concentration. In this work, using finite-element simulation, we show that the minority diffusion length can also be quantified under high optical excitation. Not only being applicable both doped and intrinsic semiconductors, the method also lifted the restriction of implementing a Schottky contact in testing devices—a condition assumed to be necessary in previous studies. The results significantly expanded the versatility of SPCM in studying a broad spectrum of semiconducting materials with unprecedented flexibility in experimental conditions.

**Introduction:**

Minority carrier diffusion length is a measure of charge carrier transport characteristics in semiconductors. It characterizes the average distance that photoexcited minority carriers can move before they recombine with majority carriers. This physical property is particulalry important in optoelectronic devices such as solar cells and photodetectors, where photoexcited charge carriers need to travel far enough in order to be collected by electrodes and contribute to the photocurrent. Longer diffusion lengths are, therefore, essential for higher quantum and energy efficiency in these devices.

Scanning photocurrent microscopy (SPCM) provided a straightforward approach for measuring the minority carrier diffusion length[1-24]. Microscopic photocurrent maps were generated by raster-scanning a tightly focused laser beam across areas of planar devices (Fig. 1a). How the charge carriers were transported across the semiconductor region could be deduced based on spatial distributions of the photocurrent. There were limitations though. Measurements reported previously in this regard were mostly conducted under two conditions: (a) photogenerated charge carriers were controlled at a low concentration using fairly weak optical intensity, the so-called low-injection condition. Negligible electric fields were induced outside of the excitation spot, minimizing the drift current (driven by electric fields) and allowing carrier diffusion to dominate the charge transport;[5, 9] (b) A depletion region due to either a Schottky or p-n junction was often present in the scanning area.[25-26] This region with a large built-in electrical potential facilitated minority carrier collection, excluding noticeable contributions from the majority carrier to the measured photocurrent. Under these conditions, photocurrent maps often showed exponential decays spatially, making it convenient to quantify the minority diffusion length. These requirements, in turn, limited the applicability of SPCM in characterizing intrinsic or lightly doped semiconductors. Because of the low concentration of intrinsic charge carriers therein, neither a large built-in potential nor the low-injection condition was feasible in experiments.[27]

Attempts were made to address such restrictions with limited success. For example, Cheng-Hao Chu *et. al.* derived an analytical model for determining the diffusion length in devices with both terminals being Ohmic contacts (without substantial built-in potential) using SPCM.[28] However, without efficient charge collection facilitated by built-in potentials, the photocurrent could be quite weak to be resolved, affecting the accuracy of quantitative analysis of photocurrent maps. Meanwhile, intrinsic semiconductors were still excluded because low optical injection remained as a condition in this analytical model. Here, using mumerical studies based on finite-element methods, we showed that SPCM could be generally applied in characterizing semiconductors, regardless of the carrier concentration or the type of electrical contacts (Ohmic or Schottky). We found that under the high-optical-injection condition, photocurrent maps would always display exponential decays. When the optical excitation intensity was sufficiently high, the photocurrent profile approached the spatial distribution of the minority carriers, allowing the minority carrier diffusion length to be measured using a simple single-exponential fitting.

**Results and discussion:**

Minority carrier diffusion length ($L_D$) in previous research was typically measured in two-terminal devices with at least one Schottky contact (Fig. 1b, inset). Upon focused laser illumination, both majority and minority carriers were generated locally and diffused due to the gradient of photogenerated charge carriers (or photocarriers in short). These carriers could also be subject to drift current when electrical fields were profound. Therefore, the photocurrent can consist of four components—diffusion and drift currents of majority and minority carriers, respectively.

Charge carrier transport, as a function of the illumination location, were simulated using finite-element methods (Comsol Multiphysics®). Material properties typical of n-doped halide perovskites with a moderate doping concentration of $N_D = 10^{15}$ cm$^{-3}$ were adopted in our simulations (Table 1).[29-31] This new class of semiconductors were intensively studied for their remarkable optoelectronic performance. Under a low-injection condition (e.g. photogeneration rate $G = 10^{21}$ cm$^{-3}$s$^{-1}$ with excess carrier concentrations $\Delta p = \Delta n = 3\times10^{12}$ cm$^{-3}$ at the laser spot), the electrical potential dropped only within the depletion region of the Schottky contact (Fig. 1b). The conduction and valence bands remained flat elsewhere across the channel, indicating negligible electric fields therein. Transport of the photocarriers in the channel was, thus, dominated by diffusion processes. Near the Schottky junction, the built-in potential prevented the majority carriers from reaching the electrode while facilitating the collection of minority carriers into the electrode. Therefore, out of the four photocurrent components, the minority carrier diffusion current was dominant. Photocurrent under this condition decreased exponentially when the laser spot was moved away from the Schottky barrier (Fig. 1c, low-injection curves, $G < 10^{23}$ cm$^{-3}$s$^{-1}$), consistent with experimental SPCM results reported previously.[10, 18, 32] The minority carrier diffusion length, in this case, would be equal to the measurable photocurrent decay length.

When the optical injection was sufficiently high (G=$10^{25}$ cm$^{-3}$s$^{-1}$, $\Delta p = \Delta n = 3.7 \times 10^{16}$ cm$^{-3}$ at the laser spot, comparable to the doping concentration $N_D$), the electronic bands in the real space showed clear dips around the laser beam spot (Fig. 1d). This phenomenon could be attributed to the photo-Dember effect.[18, 33] Because of the difference between electron and hole mobilities, higher injection rate led to more space charge accumulated at the laser spot (Fig. 1e). The space charges generated noticeable electric fields across the semiconductor channel (Fig. 1f). The photocurrent was, therefore, contributed by both carrier diffusion and drift. The SPCM profile started to deviate from single-exponential decays because of the drift components (Fig. 1c). Such distortion of the photocurrent made it particularly challenging to characterize intrinsic or lightly doped semiconductors using SPCM, because moderate optical excitation needed for measurable signals would already put the materials under high-injection condition. Similar phenomena were also reported in Electron Beam Induced Current (EBIC) measurements on AlGaAs where the position of the photocurrent maximum shifted away from the Schottky junction.[34]

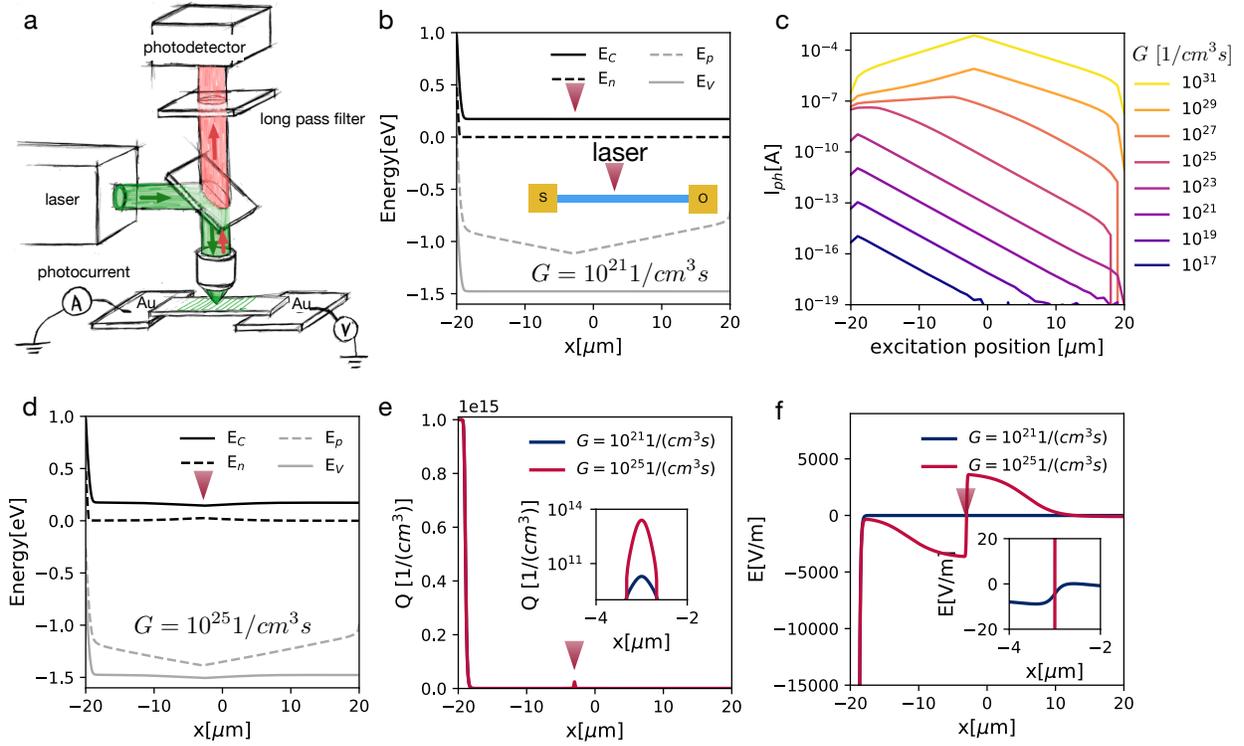

**Figure 1 Characterizing moderate-doped semiconductors using SPCM. (a)** Schematics of an SPCM setup. Besides photocurrent, additional optical signals such as photoluminescence can be mapped simultaneously with comparable spatial resolution. **(b)** Electric Potential energy of the conduction and valance bands ($E_c$, $E_v$) in an n-doped semiconductor device. The semiconductor channel is 40 mm long with Schottky (left) and ohmic contacts at the two terminals. The quasi-fermi levels of electrons and holes ($E_n$, $E_p$) are separated because of the focused laser excitation. **(c)** Photocurrent maps as a function of the laser excitation position. The profile follows a single-exponential function when the photogeneration rate G is low. **(d)** Under a higher photogeneration rate, the energy of electronic bands show dips near the laser spot, indicating a finite electric field there. **(e)** Net (space) charges Q as a result of the optical excitation were simulated under localized optical excitation. Significantly larger amount of net charges can be observed when the photogeneration rate is high. **(e)** The electric field E as a result of the net charges is stronger and uneven around the laser spot under the high-injection condition. Inset: Zoom-in around the laser spot.

| Bandgap | 1.65 eV |
|---|---|
| Electron affinity | 4.36 eV |
| Relative permittivity | 13.1 |
| Effective mass, electron | $0.1 m_e$ |
| Effective mass, hole | $0.1 m_e$ |
| Lifetime, SRH | 15 ns |
| Mobility, electron | 300 cm$^2$/(V s) |
| Mobility, hole | 100 cm$^2$/(V s) |
| Doping $N_D$ | $10^{15}$ 1/cm$^3$ |
| Schottky barrier height, $\Phi_B$ | 0.5 V |

**Table 1** The material parameters used in the simulation

We found that measuring the minority carrier diffusion length for low-doping semiconductors was not impossible even if the optical injection level became high. The possibility could be demonstrated in a semiconductor with a low doping concentration $N_D = 10^{10}$ cm$^{-3}$ (Fig. 2). The two device terminals remained as Schottky and ohmic contacts, respectively. In sharp contrast to the SPCM profiles shown in Figure 1, photocurrent maps obtained with the low-doping semiconductor did not exhibit obvious exponential decays when the optical excitation was weak (Fig. 2a, $G < 10^{24}$ cm$^{-3}$s$^{-1}$). As G was increased, a photocurrent maximum was developed in the middle of the channel, and exponential decays started to recover on both sides of the maximum. The decay length $L_{ph}$ under high optical injection approached the decay lengths of the charge carrier distributions $L_0$ (Fig. 2b). $L_0$ was a result of charge carrier transport processes (diffusion and/or drift) which, therefore, could be used to deduce the minority carrier diffusion length $L_D$. Note that such photocurrent maxima in the middle of the semiconductor channels had been observed in previous work and were ascribed to the lack of Schottky contacts (i.e. substantial built-in potentials) in the devices.[35-36] We demonstrated here when the doping concentration was low, the depletion width ($w = \sqrt{\frac{2\epsilon}{qN_D}\Phi_B}$) could easily reach or exceed the typical dimensions of electrical testing devices. The electrical potential under the so-called full- depletion condition would then be distributed evenly across the entire semiconductor channel, regardless how the electron bands were aligned at the metal (electrode)-semiconductor junctions (Fig. 2c). In other words, low-doping semidoncutors with Schottky or ohmic contacts would share similarities in SPCM measurements.

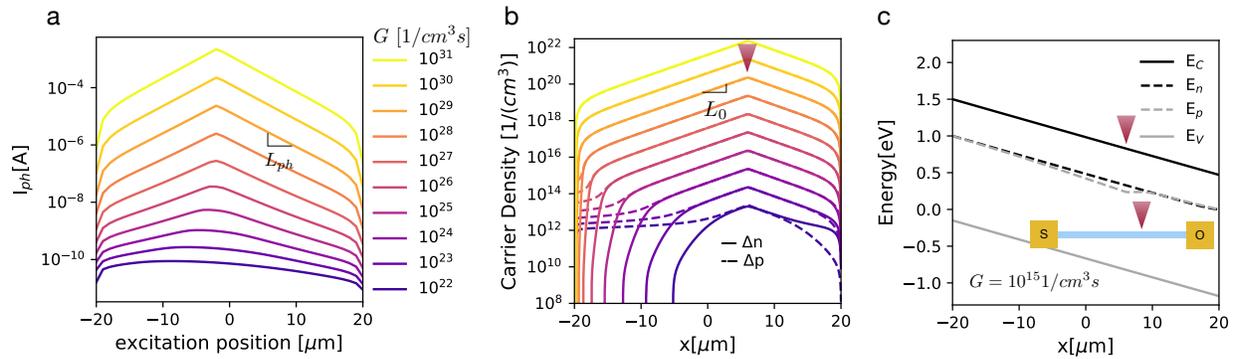

**Figure 2 Lightly-doped semiconductor device with one Schottky contact. (a)** The shape of the SPCM map changes as the photogeneration rage G increases. An electrical bias of 1 V is applied between the two terminals. **(b)** The density of excess electrons and holes as a result of focused laser excitation decays exponentially away from the excitation spot (locatated at the red triangle). **(c)** The energy of electronic bands shows a full-depletion scenario under the 1 V bias even if the left contact is configured to be Schottky.

What could make the high-injection SPCM more versatile was its applicability to devices without Schottky contacts. This was often true in experiments, where well-defined Schottky contacts were less likely to form against semiconductors with low carrier concentrations (shown numerically in Fig. 2c). We found that the spatially resolved photocurrent maps would still exhibit exponential decays even if the devices were constructed with ohmic contacts at both terminals. The decay length of the photocurrent profile could be used to deduce the minority carrier diffusion length.

The versatility of this method was demonstrated numerically using a lightly doped n-type semiconductor (Fig. 3). Both terminals of the device were configured to be ohmic contacts. Consistent with previous experimental reports[35-36], the simulated SPCM profiles showed photocurrent maxima in the middle of the semiconductor channel (Fig. 3a). Similar to the SPCM obtained with a Schottky contact (Fig. 2), exponential decays were developed on both sides of the photocurrent maxima as G was increased. This phenomenon verified that in characterizing low-carrier-concentration semiconductors, the type of electrical contacts could have less effect on the outcome of the measurements. The decay length decreased and approached a constant value $L_0$ when G was sufficiently high (Fig. 3b). $L_0$ coincided with the excess carrier decay length measured from the simulated carrier distribution, which appeared to be independent of the generation rate G within the high-injection regime (Fig 3c, $G > 10^{25}$ $cm^{-3}s^{-1}$).

In the ohmic-ohmic device, considerable charge carriers were accumulated at the location of the focused illumination, which can be attributed to the Photo Dember effect[18, 33]. The higher electron mobility assumed in the simulation resulted in a larger number of photogenerated electrons diffused away from the excitation spot, leaving behind localized positive charges (Fig. 3d). Outside of the laser spot, photogenerated injected electrons and holes follow the same exponentially decay because of the Coulomb charge screening effect[37-38], making those regions almost electrically neutral. With the G increased, more charge carriers were generated and diffused from the excitation spot to the electrodes, further broadening the charge neural regions towards the electrodes. As a result, the electric field caused by the space charges became more uniform under higher optical injection (Fig. 3e).

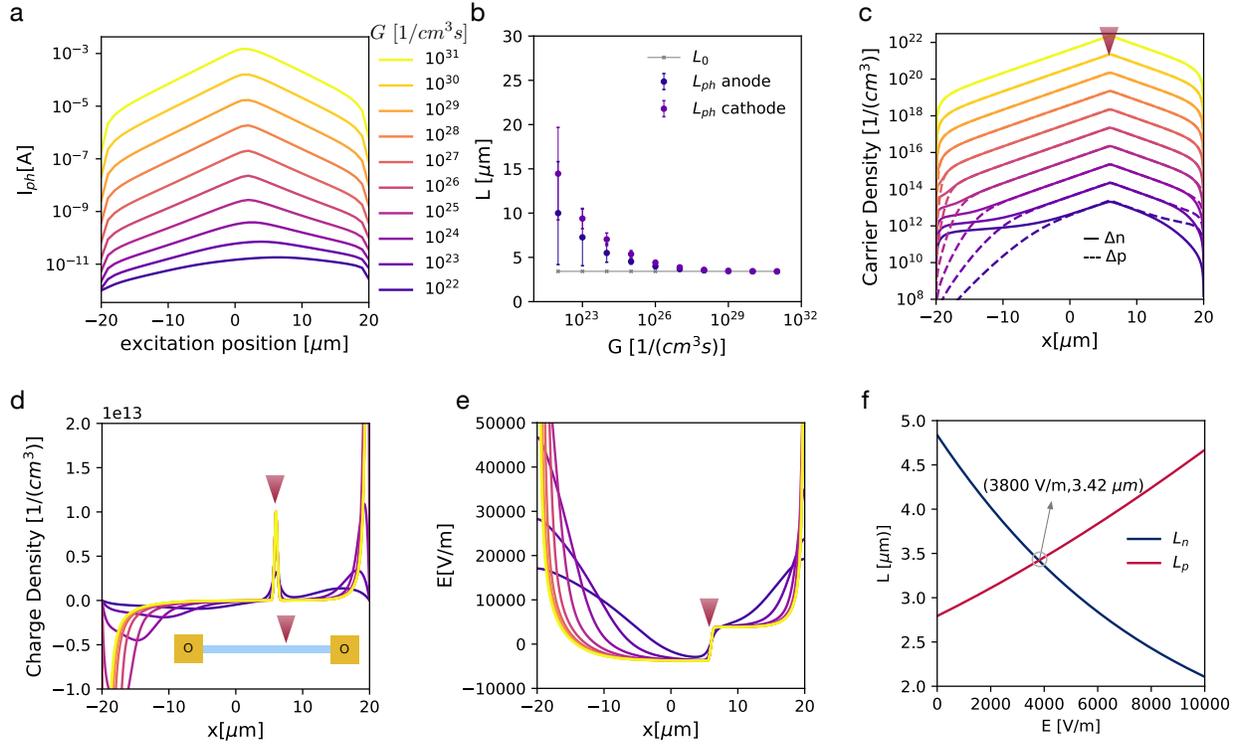

**Figure 3 Lightly-doped semiconductor device with two ohimc contacts. (a)** SPCM maps were simulated under various photogeneration rate with a moderate 0.5 V bias. **(b)** The decay length fitted from the SPCM maps ($L_{ph}$) approaches the decay length a constant value $L_0$ when G increases. **(c)** The distribution of excess carriers decays exponentially away from the laser excitation spot. The slop (exponent) appears independent from the photogeneration rate G. **(d)** The net charges ($Q = \Delta p - \Delta n$) under focused laser excitation are concentrated at laser spot and near the electrodes. **(e)** The electric field resulted from the net (space) charges become more uniform across the semiconductor channel as the photogeneration rate increases. **(f)** Plots of $L_n$ and $L_p$ as a function of electric field E, based on Eq. 7&8 yield a solution to Eq. 9 at the intersection.

The magnitude of the electric field under high optical injection was found to be independent from the injection rate G or the applied electrical biases. Instead, it was solely determined by the material characteristics including the electron and hole mobilities and the charge carrier lifetimes. Transport of both types of the charge carriers would have to follow the continuity equations:

$$\frac{1}{q}\frac{\partial J_n}{\partial x} + G - R = 0 \text{ (Eq. 1, for electrons)}$$

$$-\frac{1}{q}\frac{\partial J_p}{\partial x} + G - R = 0 \text{ (Eq. 2, for holes)}$$

The electron and hole currents could be expressed as:

$$J_n = q\mu_n nE + qD_n \frac{\partial n}{\partial x} \text{ (Eq. 3)}$$

$$J_p = q\mu_p pE - qD_p \frac{\partial p}{\partial x} \text{ (Eq. 4)}$$

Assuming the carrier capture cross sections for both types of the charge carriers were the same, the trap assisted recombination rate

$$R = \frac{pn - n_i^2}{p + n + 2n_i}\frac{1}{\tau} \text{ (Eq. 5)}$$

which can be further simplified under the high-injection condition ($\Delta n = \Delta p \gg n_0$):

$$R = \frac{\Delta n}{2\tau} = \frac{\Delta p}{2\tau} \text{ (Eq. 6)}$$

Using the above equations, we could derive the decay lengths of electrons and holes, respectively (more details in SI)

$$L_n = \frac{\sqrt{\mu_n^2 E^2 (2\tau)^2 + 4D_n(2\tau)} - \mu_n E(2\tau)}{2} \text{ (Eq. 7)}$$

$$L_p = \frac{\sqrt{\mu_p^2 E^2 (2\tau)^2 + 4D_p(2\tau)} + \mu_p E(2\tau)}{2} \text{ (Eq. 8)}$$

Because of the Coulomb charge screening effect[37-38], the decay lengths of electron and hole must be equal

$$L_p = L_n \text{ (Eq. 9)}$$

Although it could not be solved analytically, the electric field near the illumination spot can be expressed as (more details in SI)

$$E = \frac{k_B T}{q L_0} \frac{\mu_n - \mu_p}{\mu_n + \mu_p} \text{ (Eq. 10)}$$

Apparently, $E$ depended only upon the carrier mobilities and lifetimes, but not an explicit function of the generation rate G or the applied electrical biases.

The analytical derivation agreed remarkably well with results of the numerical simulation. First of all, the Coulomb charge screening effect was verified in the simulated distributions of charge carriers (Fig. 3c). The concentrations of excess electrons and holes, represented by the solid and dashed curves, respectively, overlapped with each other around the illumination spot. Furthermore, the derived delay lengths of electrons and holes ($L_n$, $L_p$) were within 1% discrepency compared to the simulated ones. $L_p$ and $L_n$, according to Eq. 7 & 8, were plotted as a function of the electric field $E$ (Fig. 3f). The condition of $L_p = L_n$ (Eq. 9) was fulfilled when the two curves intersected, where the electric field $E$ = 3800 V/m and the carrier density delay length $L_0 = L_p = L_n$ = 3.42 µm. In the numerical simulation, these quantities were found to be 3780 V/m and 3.42 µm, respectively, remarkably close to the analytical results (Fig. 3c and e).

The minority carrier diffusion length $L_D$ was found to fall in the range ($\frac{L_0}{\sqrt{2}\varphi}, \frac{L_0}{\sqrt{2}}$), where $\varphi$ happened to be the golden ratio $\frac{\sqrt{5}+1}{2}$ = 1.618. The decay length of the photocurrent ($L_{ph}$) approached the decay length of the excess carrier concentration ($L_0$) under sufficiently high optical illumination

$$L_{ph} = L_0 = \frac{\sqrt{\mu_p^2 E^2 (2\tau)^2 + 4D_p(2\tau)} + \mu_p E(2\tau)}{2} \text{ (Eq. 11)}$$

which was a function of both the minority carrier diffusion length $L_D = \sqrt{D_p \tau}$ and the drift length $L_E = \mu_p E \tau$. In experiments, the difference between the physical property $L_D$ and the measurable quantity $L_{ph}$ can be estimated if $E$ was known. Even if a precise measurement of $E$ could be challenging, the lower and upper limits could be identified by assuming $\frac{\mu_n}{\mu_p} = 1$ and $\infty$, respectively. Based on Eq. 10, plugging the range of $E$ $(0, \frac{k_B T}{\sqrt{2} q L_D})$ in Eq. 11, we obtained $(\frac{L_0}{\sqrt{2}\varphi}, \frac{L_0}{\sqrt{2}})$ to be the limits of the minority carrier diffusion length. If the ratio beteen the electron and hole mobilities could be estimated (using first-principle calculation for instance), the range of the estimated minority diffusion length could be further narrowed. It is worth noting that a decent estimate of $L_D$ could still be obtained even if a very high optical injection could not be implemented feasibly in experiments. This was because the measurable quantity $L_{ph}$ approached the saturated value rapidly as the injection rate departed from the low-injection regime. As shown in Fig. 3b, with an injection rate of $10^{23}$ cm$^{-3}$s$^{-1}$, 8 orders of magnitude lower than the highest G simulated, $L_{ph}$ was already within the same order of magnitude of $L_0$.

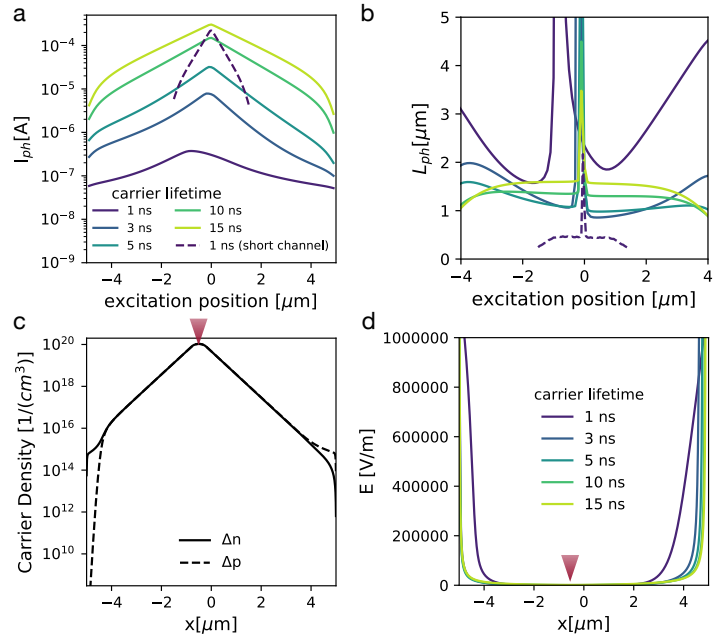

**Figure 4 The profile of SPCM maps can be modulated by varying the semiconductor channel length. (a)** Simulated photocurrent maps with the minority carrier diffusion length varied by adjusting the carrier lifetime. **(b)** The photocurrent decay length can be determined by the first-order derivative (slope) of the photocurrent maps on logarithm scales. **(c)** Net (space) charges can be clearly observed near the electrodes when the size of the semiconductor channel is much larger than the minority diffusion length (corresponding to a carrier lifetime of 1s). **(d)** Simulated distribution of electric fields shows wider regions of finite electric fields near the electrodes as the carrier lifetime decreases.

It is worth noting that the dimension of devices could compromise the feasibility of the quantitative data analysis described above. The exponential profile of the spatially resolved photocurrent would only be apparent when the minority carrier diffusion length was sufficiently long and the length of the semiconductor channel was sufficiently short.[5, 36] Alternatively, when the device became too big, the decaying profile would be distorted[35] due to insufficient diffusion and collection of minority carriers by the electrodes. Such effect could be observed in the simulated photocurrent maps when the minority carrier diffusion length was manipulated by varying the carrier lifetime (Fig. 4). Under the same high-injection condition ($G = 10^{29}$ cm$^{-3}$s$^{-1}$), the exponential decays in the photocurrent profile disappeared gradually as the carrier lifetime was reduced from 15 ns to 1 ns (Fig. 4a, corresponding to the minority diffusion length of 1.08 µm and 0.28 µm, respectively). Accordingly, a stable single-exponential fitting could be obtained using a large section of the photocurrent map when the carrier lifetime was long, whereas the derivative of the photocurrent profiles (on logarithm scales) fluctuated more with shorter carrier lifetime (Fig. 4b).

The distortion in the SPCM profiles was a result of the mismatch between the carrier diffusion length and the device channel width. When the diffusion length was too short for the photocarriers to reach the electrodes, larger differences were observed between the numbers of excess electrons and holes near both electrodes (Fig. 4c). These space charges induced electrical fields were localized near the electrodes first, and slowly extended towards the center of the device when the carrier diffusion length became short (corresponding to shorter carrier lifetime in Fig. 4d). It is the uneven electrical fields that prevented the charge transport from being predominately diffusive, distorted the single-expoential profile of the photocurrent map, and made it challenging to quantify the minority carrier diffusion length in this situation.

This problem could be simply circumvented in experiments if the size of the device could be reduced conveniently. By reducing the device size from 10 µm to 3 µm while maintaining a short carrier lifetime of 1 ns, the single-exponential decays were clearly restored in the simulated photocurrent (Fig. 4a & 4b, dashed curves). Notably, the difference between the minority carrier diffusion lengths (1.08 µm vs. 0.28 µm), dictated by the carrier lifetime (15ns vs. 1ns), was consistent with the ratio of $L_{ph}$ measured by exponential fitting (1.58 µm vs. 0.43 µm).

**Conclusions:**

Our work significantly expanded the applicability of SPCM in studying semiconducting materials. Although being widely used for more than a decade, SPCM in previous work was primarily successful on heavily doped semiconductors because of the low-injection requirement. Our results eliminated such limitation by showing the accuracy of quantitative measurements under high-injection conditions. Characterizations of heavily-doped, light-doped or even intrinsic semiconductors will all be possible. The flexbility is also extended to device fabrication.

Measurements of the minority carrier diffusion length can now be carried out in devices regardless of the types of electrical contacts, whereas at least one Schottky junction was required in previous studies. For the sake of completion, we also simulated SPCM responses of heavily-doped semiconductors under high-injection conditions—a scenario that was particularly avoided in previous experimental work (Supporting Information). The results showed that an accurate measurement of the minority diffusion length is also possible as long as the optical excitation intensity stays out of the low-injection-to-high-injection transition.

## ASSOCIATED CONTENT

**Supporting Information**

## AUTHOR INFORMATION

**Corresponding Author:** Hanwei Gao

E-mail: hanwei.gao@fsu.edu

**Author Contributions**

The research plans were conceived by both authors. X. L. carried out the finite-element simulation. Bother authors analyzed the results and wrote the manuscript.

**Funding Sources**

## ACKNOWLEDGMENT

The authors thank Dr. Xiong Peng, Dr. Kuan-Wen Chen and Mr. Zihan Zhang for insightful discussions. This work was financially supported by Office of Naval Research (N00014-18-1-2408), National Science Foundation (1930809), and the Florida State University through the Energy and Materials Initiative.